\documentclass[conference]{IEEEtran}
\IEEEoverridecommandlockouts
\usepackage{cite}
\usepackage{amsmath,amssymb,amsfonts}
\usepackage{algorithmic}
\usepackage{graphicx}
\graphicspath{ {figures/} }
\usepackage{textcomp}

\usepackage{listings}
\usepackage{array}
\usepackage{color}
\usepackage{soul}

\newcolumntype{C}[1]{>{\centering\arraybackslash}p{#1}}
\newcolumntype{D}[1]{m{#1}}

\def\BibTeX{{\rm B\kern-.05em{\sc i\kern-.025em b}\kern-.08em
    T\kern-.1667em\lower.7ex\hbox{E}\kern-.125emX}}
\begin{document}

\bstctlcite{IEEEexample:BSTcontrol}

\title{Which Source Code Plagiarism Detection Approach is More Humane?
}

\author{\IEEEauthorblockN{Oscar Karnalim}
\IEEEauthorblockA{\textit{Faculty of Information Technology} \\
\textit{Maranatha Christian University}\\
Bandung, Indonesia \\
oscar.karnalim@it.maranatha.edu}
\and 
\IEEEauthorblockN{Lisan Sulistiani}
\IEEEauthorblockA{\textit{Faculty of Information Technology} \\
\textit{Maranatha Christian University}\\
Bandung, Indonesia \\
lisans1601@gmail.com}
}

\maketitle

\begin{abstract}
This paper contributes in developing source code plagiarism detection that is more aligned with human perspective.
Three evaluation mechanisms that directly relate human perspective with evaluated approaches are proposed: think-aloud, aspect-oriented, and empirical mechanism.
Using those mechanisms, a comparative study toward attribute- and structure-based plagiarism detection approach (i.e., two popular approach categories in source code plagiarism detection) is conducted.
According to that study, structure-based approach is more effective than the attribute-based one;
its signature aspect and resulted similarity degrees are more related to human preferences.
In addition, such approach is related to most human-oriented aspects for suspecting source code plagiarism.
\newline
\end{abstract}

\section{INTRODUCTION}
Source code plagiarism occurs when a student claims another student's source code as their original work \cite{Cosma2008}.
In programming courses, it is considered as an illegal behavior;
the correlation between student's knowledge and grade will be weakened, resulting more complicated recruitment process for industry. 
Mitigating such illegal behavior is a crucial task for programming examiners (e.g., lecturers or assistant lecturers).

To mitigate human effort, automated plagiarism detection approaches have been developed \cite{Lancaster2004}.
Using such approaches, plagiarism-suspected pairs could be detected in a no time.
However, most approaches (where some of them are implemented on publicly available tools such as JPlag \cite{Prechelt2002}) could be either too sensitive or insensitive when compared to human.
Some of them could generate numerous false positives (i.e., non-plagiarized pairs which are considered as plagiarism-suspected pairs) while the others could generate numerous false negatives (i.e., plagiarized pairs which are not considered as plagiarism-suspected pairs).
The existence of false results would complicate accusation process: examiner needs to double-check each result comprehensively to avoid mis-detection.

One of the solutions for reducing false results is to utilize a plagiarism detection approach which behaves like humans (i.e., examiners), considering correct detection results are originally defined by them.
Therefore, it is natural to evaluate plagiarism detection approaches from human perspective.
Nevertheless, to the best of our knowledge, existing evaluation mechanisms only indirectly consider human perspective. 
We would argue that direct relation between human perspective and evaluated plagiarism detection approaches are needed to favor human-like approach in evaluation.

This paper proposes two contributions.
First, three evaluation mechanisms that directly relate human perspective to evaluated plagiarism detection approaches are proposed.
Such mechanisms could become evaluation alternatives in addition to conventional mechanism (which relies on human-annotated dataset).
Second, using proposed evaluation mechanisms, a comparative study between attribute- and structure-based approach (i.e., two well-known approach categories in plagiarism detection \cite{Al-Khanjari2010,Karnalim2017IAENG}) is presented.
This study draws the applicability of proposed evaluation mechanisms.

\section{RELATED WORKS}
In general, source code plagiarism detection approaches work in two phases: conversion and comparison phase.
Conversion phase translates given source codes and/or their features to intermediate representations for more effective and efficient comparison phase.
Some of frequently-used intermediate representations are: source code token sequence \cite{Prechelt2002,Engels2007,ElBachirMenai2010,Ottenstein1976,FaidhiJ.A.W;Robinson1987,Acampora2015,Inoue2012,Ohmann2015,Mozgovoy2005,Moussiades2005,Al-Khanjari2010,Burrows2007,Poon2012,Duric2013,Bejarano2015,Kustanto2009}, source code word sequence (i.e., a sequence generated from natural language tokenizer) \cite{flores2015cross,Cosma2012}, low-level token sequence (i.e., a sequence generated from executable file of given source code) \cite{Karnalim2016,Rabbani2017,Karnalim2017IAENG,Karnalim2017ICSESS,Karnalim2018}, program dependency graph \cite{Liu2006}, and abstract syntax tree\cite{Fu2017,Ganguly2017}.
After translated, resulted intermediate representations are then passed to comparison phase---that measures similarity degree and defines which pairs are suspected as plagiarism cases. 
With regard to this phase, plagiarism detection approaches can be further classified into three categories: attribute-based, structure-based, and hybrid approach \cite{Al-Khanjari2010,Karnalim2017IAENG}.

Attribute-based approach determines plagiarism-suspected pairs based on shared source code characteristics (e.g., the number of unique operators \cite{Ottenstein1976} and token occurrences \cite{Acampora2015}).
Initially, similarity measurement on such approach relies on attribute-counting mechanism;
two codes are considered similar if both of them share the same characteristic occurrence frequencies \cite{Ottenstein1976,FaidhiJ.A.W;Robinson1987,Vamplew2010}. 
Later, advanced similarity measurements are incorporated.
These measurements are adapted from Fuzzy Logic (e.g., Fuzzy C-Means \cite{Acampora2015}), Machine Learning (e.g., Random Forest \cite{Ganguly2017}), and Information Retrieval (e.g., Cosine Similarity \cite{Inoue2012,Ohmann2015}, Jaccard Coefficient \cite{Moussiades2005}, Language Model \cite{Ganguly2017}, and Latent Semantic Analysis \cite{Cosma2012,flores2015cross}).

Structure-based approach determines plagiarism-suspected pairs based on shared source code structure.
This approach typically relies on string-matching algorithms---such as Running-Karp-Rabin Greedy-String-Tiling \cite{Prechelt2002,Kustanto2009,Karnalim2016,Karnalim2017ICSESS,Karnalim2017IAENG,Karnalim2018,Bejarano2015,Duric2013}, local alignment \cite{Rabbani2017}, and Winowing algorithm \cite{Duric2013}---that have been modified to handle token streams.
However, considering the limitation of those algorithms, some graph-matching algorithms (e.g., tree kernel algorithm \cite{Fu2017} and graph isomorphism algorithm \cite{Liu2006}) are proposed as alternative comparison algorithms.

Hybrid approach combines both attribute- and structure-based approach for determining plagiarism-suspected pairs.
Such combination aims to enhance either effectiveness \cite{Poon2012,Engels2007,ElBachirMenai2010} or efficiency \cite{Burrows2007}. 
For enhancing effectiveness, these combined approaches either show the results of both conventional approaches at once \cite{ElBachirMenai2010} or treat the result of one approach as an input of another (e.g., using the result of structure-based approach as an attribute for learning algorithm \cite{Engels2007} or clustering algorithm \cite{Poon2012}).
On the other, for enhancing efficiency, plagiarism-suspected pairs from attribute-based approach are used as the inputs of structure-based approach \cite{Burrows2007}. 
In such manner, not all source pairs are compared using structure-based approach (which is typically costly in terms of processing time).

In the context of effectiveness evaluation, two mechanisms are frequently used in source code plagiarism detection: statistical and domain-specific mechanism.
Statistical mechanism utilizes metrics commonly used in statistic (such as precision) to determine the performance of given plagiarism detection approaches \cite{Acampora2015,Kustanto2009,flores2015cross,Inoue2012,Ohmann2015,Burrows2007,Prechelt2002,Moussiades2005,Al-Khanjari2010,Engels2007,Ganguly2017,Fu2017,Cosma2012,Duric2013,Mozgovoy2005}.
This mechanism requires an evaluation dataset covering all statistical result categories (i.e., true positive, true negative, false positive, and false negative).
In contrast, domain-specific mechanism utilizes metrics specifically designed for plagiarism detection (e.g., similarity degree) to determine plagiarism detection approaches' performance \cite{Rabbani2017,Karnalim2016,Karnalim2017IAENG,Karnalim2018Icoict,Karnalim2017ICSESS,Karnalim2018,Bejarano2015,Liu2006,Poon2012,ElBachirMenai2010}.
Unique to this mechanism, it only relies on true positives (that are frequently created in controlled environment).
In other words, it is more practical to be used (since only true positives are required) yet it is less related to real environment (since other result categories are not considered).

One of the objectives in developing automated source code plagiarism detection is to let computer imitates manual plagiarism detection done by human.
Hence, it is natural to evaluate the effectiveness of plagiarism detection approaches from human perspective.
However, existing evaluation mechanisms (i.e., statistical and domain-specific mechanism) may not sufficiently reflect human perspective; considering such perspective is not directly related to evaluated approaches; 
they are only related in indirect manner through human-annotated dataset.
Such dissonance could complicate accusation process due to the existence of false results.
\section{HUMAN-ORIENTED EVALUATION MECHANISMS}
Three human-oriented evaluation mechanisms (i.e., evaluation mechanisms that directly relate human perspective with evaluated plagiarism detection approaches) are proposed: think-aloud, aspect-oriented, and empirical evaluation mechanism. 
To mitigate qualitative bias, human-oriented effectiveness is determined by comparing plagiarism detection approaches to each other through questionnaire survey.
Considering such comparison is best performed with only two instances, proposed evaluation mechanisms are designed to accept two approaches per execution.
However, if necessary, they could compare numerous approaches by adapting tournament selection from Genetic Algorithm \cite{russell2003artificial} (with an assumption that each approach refers to one instance and fitness function is the human-oriented effectiveness degree).

\subsection{Think-Aloud Evaluation Mechanism}
Think-aloud evaluation mechanism compares the effectiveness of two source code plagiarism detection approaches based on respondents' descriptions about how manual detection works.
It is suitable to be used when only a limited number of plagiarism cases exists for evaluation and the main goal is to observe from process perspective (i.e., factors considered while detecting plagiarism).
This mechanism is inspired from think-aloud protocols used in Empirical Software Engineering \cite{Singer2008}. 

Think-aloud evaluation mechanism consists of three phases.
First of all, respondents are asked to manually suspect plagiarized source codes for reminding them how manual detection works.
Second, they should describe how they suspected plagiarized codes as detail as possible in natural language sentences.
Third, each aspect from resulted descriptions will be qualitatively linked to evaluated plagiarism detection approaches; 
an approach is more effective than another if its characteristics are frequently mentioned on respondents' descriptions.
To mitigate bias, no clue should be given to respondents in regard to evaluated plagiarism detection approaches.

It is important to note that this mechanism might not capture some human-oriented aspects due to human limitations namely low interpretation skill and awareness.
Low interpretation skill occurs when the respondent cannot formalize what is in their mind while suspecting plagiarized source codes;
whereas, low awareness occurs when the respondent does not aware about some parts of detection process (which have been unconsciously conducted). 
\subsection{Aspect-Oriented Evaluation Mechanism}
Aspect-oriented evaluation mechanism relies on one differentiating aspect between two source code plagiarism detection approaches to measure human-oriented effectiveness.
If that aspect affects (or positively correlates more with) humans' preferences, it can be stated that an approach with such aspect is more effective.
Otherwise, the approach without such aspect is preferred. 
This mechanism is suitable to be used when there is a salient differentiating aspect between both plagiarism detection approaches and the main goal is to observe from result perspective (i.e., resulted similarity degree).
In general, this mechanism works by asking respondents to rank plagiarism cases from human perspective wherein the results will be further analyzed in qualitative and quantitative perspective.
It is inspired from ranking concept in Information Retrieval \cite{Croft2010} where relevancy (i.e., similarity in our case) is defined based on ranks. 

Prior to asking respondents, survey cases (that accentuate the impact of differentiating aspect) should be artificially developed. 
For each case, an original source code is plagiarized to several codes where the only modification is the differentiating aspect with various degree.
To guarantee that the impact of differentiating aspect is shown, on such cases, the difference between both evaluated approaches in terms of similarity degree should be statistically significant.
The significance itself can be measured using t-test.

For each survey case, respondents are asked to rank plagiarized codes in descending order based on their similarity degree toward original code from human perspective.
Each plagiarized code will be assigned with a rank where the first rank refers to the highest similarity;
if two or more codes are equally similar, they could be assigned with similar rank.
To provide more objective results, each survey case should be rated by more than one respondent.

After all survey cases are ranked by the respondents, human-oriented effectiveness of evaluated approaches can be analyzed from both qualitative and quantitative perspective.
From qualitative perspective, an approach with aforementioned differentiating aspect is more effective if human-assigned ranks are changed as the degree of involved differentiating aspect is modified. 
In contrast, from quantitative perspective, the effectiveness of evaluated plagiarism detection approaches is defined through Pearson correlation between similarity degrees and averaged human-assigned ranks; 
a plagiarism detection approach is more effective than another if its resulted similarity degrees (measured on artificially-plagiarized codes toward their originals) generate higher correlation toward averaged human-assigned ranks.
Considering similarity degree's result interpretation (where higher value is preferred) contradicts human-assigned rank's (where lower value is preferred), averaged human-assigned ranks will be negated before correlating to make its result interpretation proportional to similarity degree's.

\subsection{Empirical Evaluation Mechanism}
Empirical evaluation mechanism relies on humans' preferences toward contradicting plagiarism pairs (i.e., pairs where each of their members exclusively favors a plagiarism detection approach), selected from existing plagiarism dataset, to evaluate human-oriented effectiveness.
This mechanism is suitable to be used when the main goal is to observe from result perspective (i.e., resulted similarity degree) and a publicly-available plagiarism dataset is involved.
It is inspired from ranking concept in Information Retrieval \cite{Croft2010} (where relevancy is defined based on ranks) and A/B testing in Web Analytics\cite{Kohavi2017} (where human should select one between two instances based on their qualitative perspective).

Contradicting plagiarism pairs are selected from a plagiarism dataset in fourfold.
First, plagiarized codes are clustered based on their original code (one cluster for each original code).
Second, per plagiarism detection approach, plagiarized codes will be compared with their originals, resulting a set of similarity degrees.
Third, for each cluster and a plagiarism detection approach, plagiarized codes are ranked in descending similarity order.
Fourth, contradicting plagiarism pairs are selected for each cluster.
Two plagiarized source codes (e.g., \textit{A} and \textit{B}) are considered as a contradicting plagiarism pair \textit{iff} returned value from either (\ref{eq:contradiction1}) or (\ref{eq:contradiction2}) is true.
\textit{P1} returns a rank resulted from the first plagiarism detection approach while \textit{P2} returns a rank resulted from another.
\begin{equation}
Con1(A, B) = (P1(A) > P1(B)) \land (P2(A) < P2(B))
\label{eq:contradiction1}
\end{equation}
\begin{equation}
Con2(A, B) = (P1(A) < P1(B)) \land (P2(A) > P2(B)) 
\label{eq:contradiction2}
\end{equation}

To provide more insight, suppose we have three plagiarized codes (\textit{A}, \textit{B}, and \textit{C});
their similarity degree toward original code---measured using two detection approaches: \textit{P1} and \textit{P2}---can be seen on Table \ref{tab:example}.
According to given example, \textit{P1}'s and \textit{P2}'s ranking order are \textit{\{A, C, B\} }and \textit{\{C, A, B\}} respectively.
As a result, \textit{A} and \textit{C} are considered as a contradicting pair; 
they satisfy (\ref{eq:contradiction1}) since \textit{P1(A)} $>$ \textit{P1(C)} and \textit{P2(A)} $<$ \textit{P2(C)}.

\begin{table}[htbp]
\caption{Similarity degrees for illustrating how contradicting plagiarism pairs are selected}
\label{tab:example}
\centering
\begin{tabular}{| C{0.075\textwidth}| C{0.15\textwidth}| C{0.15\textwidth}|}
\hline
\bfseries Plagiarized Code & \bfseries P1's Similarity Degree & \bfseries P2's Similarity Degree\\
\hline
A
& 
70\%
&
50\% \\
\hline
B
& 
50\%
&
40\% \\
\hline
C
& 
60\%
&
95\% \\
\hline
\end{tabular}
\end{table}

Two things should be considered while selecting contradicting pairs.
First, plagiarized codes in each contradicting pair should be explicitly different to each other when perceived by human. 
Hence, it is preferable to select pairs that generate high delta similarity degree between evaluated approaches.
Second, the difference between evaluated approaches should not be coincidental on selected contradicting pairs.
Hence, it is important to assure that similarity degrees resulted from both approaches are significantly different to each other; wherein the significance can be measured using t-test.

For each contradicting pair, each respondent should select one plagiarized code that seems more similar to original code;
preferred code will be assigned with the $1^{st}$ rank while another will be assigned with the $2^{nd}$ rank.

The result of this evaluation can be quantitatively analyzed by correlating averaged human-assigned ranks with resulted similarity degrees (measured on plagiarized codes with their originals per evaluated plagiarism detection approach) using Pearson correlation;
higher correlation refers to higher effectiveness degree.
Similar to aspect-oriented mechanism, averaged human-assigned ranks should be negated beforehand to make its result interpretation proportional to similarity degree's (where higher value is preferred).
Further, the effectiveness of evaluated plagiarism detection approaches can also be naively measured by counting how many favoring codes are assigned with the $1^{st}$ rank for each approach.

Qualitative analysis is discouraged to be conducted.
Plagiarized codes are not specifically designed to accentuate the difference between evaluated plagiarism detection approaches.
It will therefore be difficult to qualitatively collect findings regarding why a favoring code is preferred than another.

\section{CASE STUDY: COMPARING ATTRIBUTE- WITH STRUCTURE-BASED APPROACH}
As our case study, proposed evaluation mechanisms will be applied to compare human-oriented effectiveness of Attribute-Based Approach (ABA) and Structure-Based Approach (SBA).
These approaches are the baseline of most plagiarism detection approaches \cite{Al-Khanjari2010,Karnalim2017IAENG}.
We are aware that a work proposed in \cite{Verco1996} also compares those categories. 
Yet, human perspective is not directly linked to plagiarism detection approaches in their evaluation.

Both ABA and SBA work by accepting two source codes to return a similarity degree.
At first, inputted source codes are tokenized using ANTLR \cite{Parr2013} (with an exclusion of comment tokens).
Later, resulted token sequences will be compared to each other using specific similarity algorithm.
ABA will utilize Vector Space Model and Cosine Similarity \cite{Croft2010} (where each vector refers to token occurrence frequency).
Whereas, SBA will utilize Running-Karp-Rabin Greedy-String-Tiling algorithm \cite{Wise1995}, with two as its minimum matching threshold and average similarity equation \cite{Wise1992} as its normalization technique.

Fifteen lecturer assistants are involved in this evaluation.
They are experienced in terms of checking students' programming assignments.
Hence, it is expected that resulted evaluation findings will be considerably objective.
It is true that the number of respondents are somewhat limited.
However, since the main goal of this case study is to prove the applicability of proposed evaluation mechanism, we would argue such limited number of respondents are enough.

\subsection{Methodology}
\begin{figure}[htbp]
\centerline{\includegraphics[width=0.2\textwidth]{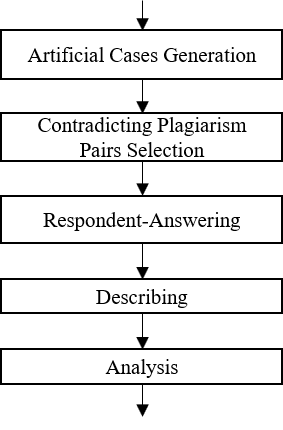}}
\caption{Combined phases from proposed evaluation mechanisms}
\label{fig:methodology}
\end{figure}

Proposed evaluation mechanisms will be merged as one in this case study.
Fig. \ref{fig:methodology} shows that such combination generates five phases: artificial cases generation, contradicting plagiarism pairs selection, respondent-answering, describing, and analysis phase.

Artificial cases (for aspect-oriented evaluation mechanism) are generated by considering token order as a differentiating aspect;
such aspect is only considered by SBA while determining similarity.
Four artificial cases are considered in this study, covering various scopes of order-changed token sequences: single-instruction, multiple-instructions, method, and class scope.
All of them are written in Java programming language.
A case with single-instruction scope has 4 plagiarized codes that are formed by swapping \textit{N} instructions (where \textit{N} in these codes are 0, 1, 3, and 5 respectively).
Remaining cases contain 6 plagiarized codes each, that are formed by changing the order of three subsequences in combinatoric manner. 
Generated artificial cases are valid to be used in our evaluation;
their SBA's and ABA's result are statistically different (in terms of similarity degree) when measured using two-tailed paired t-test. It generates p=0.02, which is lower than 0.05.


Contradicting plagiarism pairs (for empirical evaluation mechanism) are selected from a Java-based plagiarism dataset \cite{Karnalim2018Icoict} by clustering plagiarized codes per original code and plagiarism level \cite{FaidhiJ.A.W;Robinson1987}, resulting 42 clusters (plagiarism level is added to our consideration for providing more comprehensive findings).
Forty-five contradicting pairs are selected for our survey;
they are formed from level-2 to level-6 plagiarism categories where each category contributes 5, 9, 11, 10, 10 contradicting pairs respectively.
Those pairs lead to a statistically significant difference between ABA and SBA when measured using two-tailed paired t-test.
It generates p-value=1.59E-78, which is lower than 0.05.
Hence, they are valid to be used in this evaluation.

Regarding artificial cases and contradicting plagiarism pairs, respondents will be asked to rank them in similar manner as defined in aspect-oriented and empirical evaluation mechanism.
However, considering ranking 45 contradicting pairs (for empirical evaluation mechanism) at once may generate biased result due to human fatigue, our respondents are divided to three groups of five where the member of each group is only responsible to rank one-third of total pairs.

After ranking artificial cases and contradicting plagiarism pairs, describing phase will be conducted. 
At this phase, respondents will be asked to write down their manual detection technique as detail as possible.

Survey results collected from respondent-answering and describing phase will be further analyzed in qualitative and quantitative manner.
These analysis works exactly similar as in our proposed evaluation mechanisms.

\subsection{Results}
According to our evaluation, token order (which is SBA's signature aspect) is considered by respondents when detecting plagiarism cases. 
Plagiarized code without such change (i.e., a verbatim copy of original code) is always assigned with the highest rank when compared to its counterpart with such change.
Further, resulted negated ranks are strongly correlated with SBA's similarity degrees in positive manner;
their correlation degree is 0.833 when measured using Pearson correlation (where 1 represents the strongest positive correlation).
Consequently, it can be stated that SBA is preferred than ABA from aspect-oriented evaluation perspective.

It is important to note that the correlation between negated ranks and ABA's similarity degrees is immeasurable due to Pearson correlation's nature.
It cannot correlate two sequences when one of them has no variability (which is ABA in our case study, it generates resemblant similarity degrees for all artificial cases).

\begin{figure}[htbp]
\centerline{\includegraphics[width=0.45\textwidth]{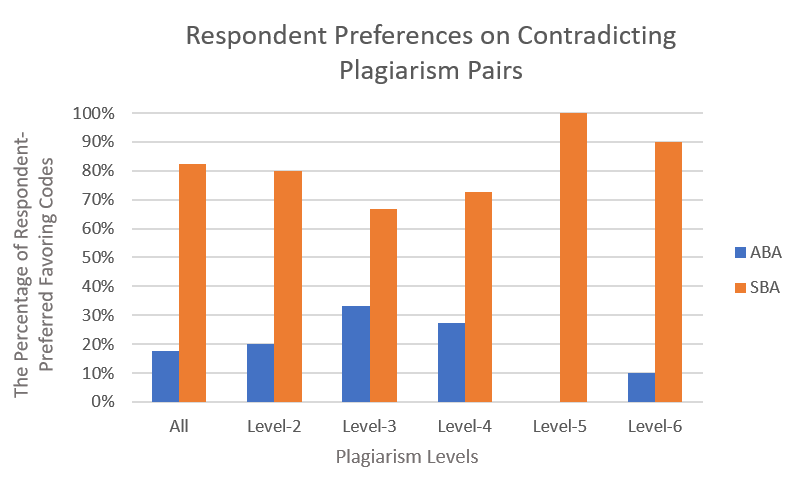}}
\caption{The percentage of respondent-preferred favoring codes on contradicting plagiarism pairs. Each number is generated by counting how many favoring codes are assigned with the $1^{st}$ rank and then normalized based on the number of involved plagiarism pairs.}
\label{fig:preference_empirical}
\end{figure}

\begin{figure}[htbp]
\centerline{\includegraphics[width=0.45\textwidth]{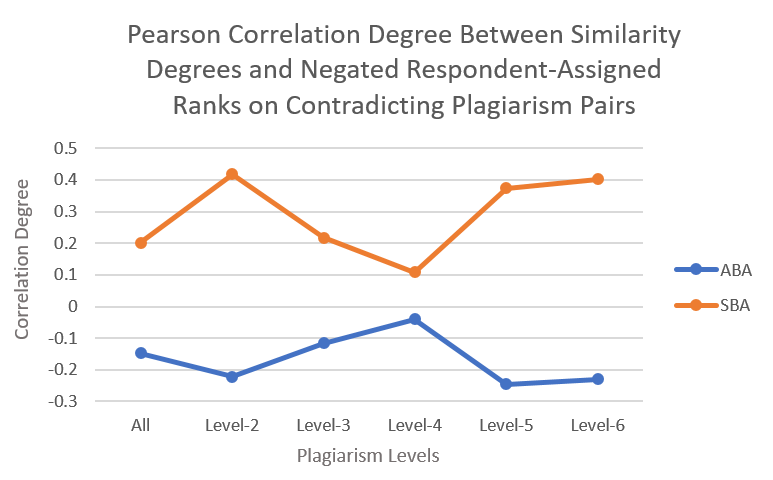}}
\caption{Pearson correlation degree between similarity degrees and negated respondent-assigned ranks on contradicting plagiarism pairs.}
\label{fig:correlation_empirical}
\end{figure}

When perceived from empirical evaluation, SBA is preferred than ABA in both general and plagiarism level perspective.
Fig. \ref{fig:preference_empirical} shows that SBA's percentage of respondent-preferred codes is far higher than ABA's.
Further, SBA's similarity degrees are more positively correlated with negated respondent-assigned ranks despite their low correlation. 
As seen in Fig. \ref{fig:correlation_empirical}, on all categories, SBA generates positive correlation while ABA generates the negative one.
Low correlation between similarity degrees and respondent-assigned ranks is natural considering plagiarized codes are not specifically designed to accentuate the difference between evaluated plagiarism detection approaches.

SBA generates the strongest correlation on level-2 plagiarism level category (which is about identifier renaming).
Hence, it can be stated that respondents tend to detect plagiarism as SBA when source code structure is unchanged.
In contrast, it generates the weakest correlation on level-4 plagiarism level category.
In other words, method structure change (i.e., level-4 signature attack) makes respondents focus less on source code structure. 

Respondents' descriptions about their manual detection technique show that six aspects are considered while suspecting plagiarism cases.
Table \ref{tab:think_aloud_considered_aspects} displays those aspects, including their occurrences (calculated based on the number of respondents mentioning it) and their relationship to evaluated plagiarism detection approaches.
Most of mentioned aspects are related to SBA.
Hence, it can be stated that SBA is more effective than ABA from think-aloud evaluation perspective.

\begin{table}[htbp]
\caption{Considered aspects while suspecting plagiarism cases}
\label{tab:think_aloud_considered_aspects}
\centering
\begin{tabular}{| D{0.08\textwidth}| C{0.07\textwidth}| D{0.26\textwidth}|}
\hline
\bfseries \hspace{0.2cm} Aspect & \bfseries Occurrences & \bfseries Relationship to Evaluated Approaches\\
\hline
Statement order
&
11
& 
Related to SBA since order is a part of source code structure.
\\
\hline
Semantic
&
5
&
Related to SBA since semantic is preserved based on source code structure.
\\
\hline
Identifier name
&
3
& 
Related to ABA since identifier name is a source code characteristic.
\\
\hline
Structure
&
2
&
Obviously related to SBA.
\\
\hline
Output
&
1
& 
Related to SBA since output is defined based on source code structure.
\\
\hline
Line of code
&
1
& 
Related to ABA since line of code is a source code characteristic.
\\
\hline
\end{tabular}
\end{table}

To sum up, from human perspective, SBA is more effective than ABA according to three rationales.
First, SBA's signature aspect (i.e., token order) is considered by respondents when detecting plagiarism cases.
Second, SBA's favoring cases is preferred than ABA's.
Third, SBA is related to more human-oriented aspects for suspecting source code plagiarism.

\section{CONCLUSION AND FUTURE WORK}
In this paper, three human-oriented evaluation mechanisms in source code plagiarism detection have been proposed.
Unique to these mechanisms, they directly relate human perspective with evaluated plagiarism detection approaches.
These mechanisms have been tested on a case study toward attribute- and structure-based plagiarism detection approach.
According to such test, structure-based approach is more effective; its signature aspect and similarity degree is more related to respondents' preferences.
Further, it is related to most aspects that are mentioned by respondents for suspecting plagiarism cases.

For future works, we plan to evaluate human-oriented effectiveness of frequently-used features on source code plagiarism detection (such as method linearization \cite{Karnalim2017ICSESS,Karnalim2018}) using aspect-oriented evaluation mechanism.
Further, we also plan to systematize how human suspects source code plagiarism (with and without machine learning algorithm).
The results of such work (e.g., features with high human-oriented effectiveness degree and systematic steps to suspect plagiarism) will be used to develop source code plagiarism detection approach that is more aligned to human perspective.




\bibliographystyle{IEEEtran}
\bibliography{IEEEabrv,references}

\end{document}